\documentclass[preprint2]{aastex}

\catcode`\@=11
\newcommand{\gapprox}{\mathrel{\mathpalette\@versim>}}
\newcommand{\lapprox}{\mathrel{\mathpalette\@versim<}}
\newcommand{\propapprox}{\mathrel{\mathpalette\@versim\propto}}
\newcommand{\@versim}[2]
  {\lower3.1truept\vbox{\baselineskip0pt\lineskip0.5truept
\ialign{$\m@th#1\hfil##\hfil$\crcr#2\crcr\sim\crcr}}}
\catcode`\@=12

\shorttitle{THE EVOLUTION OF PSR J0737$-$3039B AND A MODEL FOR RELATIVISTIC SPIN PRECESSION}
\shortauthors{PERERA ET AL.}

\begin{document}

\title{The evolution of PSR J0737$-$3039B and a model for relativistic spin precession}

\author{B.~B.~P.~Perera,\altaffilmark{1} M.~A.~McLaughlin,\altaffilmark{1,2,3} M.~Kramer,\altaffilmark{4}
 I.~H.~Stairs,\altaffilmark{5} R.~D.~Ferdman,\altaffilmark{6} P.~C.~C.~Freire,\altaffilmark{7} 
 A.~Possenti,\altaffilmark{8} R.~P.~Breton,\altaffilmark{9,10} R.~N.~Manchester,\altaffilmark{11}
 M.~Burgay,\altaffilmark{8} A.~G.~Lyne,\altaffilmark{4} $\&$ F.~Camilo\altaffilmark{12}}

\altaffiltext{1}{Department of Physics, West Virginia University, Morgantown, WV 26506, USA.}
\altaffiltext{2}{National Radio Astronomy Observatory, Green Bank, WV 24944,USA.}
\altaffiltext{3}{Alfred P. Sloan Research Fellow.}
\altaffiltext{4}{University of Manchester, Jodrell Bank Observatory, Macclesfield SK11 9DL, UK.}
\altaffiltext{5}{Department of Physics and Astronomy, University of British Columbia, 6224 Agricultural Road, Vancouver,British Columbia V6T 1Z1, Canada.}
\altaffiltext{6}{University of Manchester, School of Physics and Astronomy, Jodrell Bank Center for Astrophysics, Alan Turing Building, Manchester, M13 9PL, UK.}
\altaffiltext{7}{NAIC, Arecibo Observatory, HC03 Box 53995, PR 00612, USA.}
\altaffiltext{8}{INAF-Osservatorio Astronomica di Cagliari, Loc. Poggio dei Pini, Strada 54, 09012 Capoterra, Italy.}
\altaffiltext{9}{Department of Physics, McGill University, Montreal, QC H3A 2T8, Canada.}
\altaffiltext{10}{Department of Astronomy and Astrophysics, University of Toronto, Toronto, ON M5S 3H4, Canada.}
\altaffiltext{11}{Australia Telescope National Facility, Commonwealth Scientific and Industrial Research Organisation (CSIRO), P.O. Box 76, Epping, New South Wales 1710, Australia.}
\altaffiltext{12}{Columbia Astrophysics Laboratory, Columbia University, 550 West 120th Street, New York, NY 10027, USA.}

\vskip 1 truein

\newpage

\begin{abstract}

We present the evolution of the radio emission from the $2.8$-s pulsar of the double pulsar system PSR J0737$-$3039A/B. We provide an update on the \citet{bpm+05} analysis by describing the changes in the pulse profile and flux density over five years of observations, culminating in the B pulsar's radio disappearance in 2008 March. Over this time, the flux density decreases by $0.177$~mJy~yr$^{-1}$ at the brightest orbital phases and the pulse profile evolves from a single to a double peak, with a separation rate of $2.6$$\degr$~yr$^{-1}$.
The pulse profile changes are most likely caused by relativistic spin precession, but can not be easily explained with a circular hollow-cone beam as in the model of \citet{cw08}.
Relativistic spin precession, coupled with an elliptical  beam, can model the pulse profile evolution well and the reappearance is expected to happen in $\sim2035$ with the same part of the beam or in $\sim2014$ if we assume a symmetric beam shape. This particular beam shape predicts geometrical parameters for the two bright orbital phases which are consistent and similar to those derived by \citet{bkk+08}.
However, the observed decrease in flux over time and B's eventual disappearance cannot be easily explained by the model and may be due to the changing influence of A on B.

\end{abstract}


\section{Introduction}
\label{intro}

Double neutron stars are systems of two neutron stars orbiting one another. Only ten such systems are known so far \citep{stairs08,kkl+09}.
The detection of both neutron stars as radio pulsars is a rare situation, as it requires the electromagnetic beam from both pulsars to intercept our line-of-sight during the short interval when both pulsars are active.
The only such system, J0737$-$3039A/B, was discovered in late $2003$ with the $64$-m Parkes radio telescope in Australia as part of a high-latitude multibeam survey of the Southern sky \citep{bdp+03,lbk+04}. It provides an amazing laboratory for the study of relativistic gravity and the most precise test to date of general relativity (GR) in the strong-field regime \citep{ksm+06}. The two pulsars orbit each other in a $2.4$-hr orbit, the shortest of any observed double neutron star binaries, with moderate orbital eccentricity of 0.088.

The first-born pulsar is believed to have been recycled, hereafter referred to as `A', has a  short spin period of $23$ ms and the second-born companion, hereafter `B', spins with a longer $2.8$ s period. With their strong gravitational fields and rapid motions, the system is expected to show large relativistic effects and, indeed, has a relativistic advance of periastron of $17$$\degr$~yr$^{-1}$ \citep{ksm+06}, the highest rate of any observed pulsar binary.
Another of these effects is the relativistic spin precession, also known as geodetic precession, of the pulsar's spin axis about the orbital angular momentum. The theoretical equation formulated by \citet{bo75b} and \citet{ber+75} predicts a spin precession rate of $5.061(2)$$\degr$~yr$^{-1}$ for pulsar B\footnote{Here, and throughout the paper,
the number in parentheses is the  1-$\sigma$ uncertainty  in the last quoted digit.}. Recently this rate has been constrained to be $4.8(7)$$\degr$~yr$^{-1}$ by fitting a model to the changing morphology of the eclipses of A seen when it passes behind the absorbing magnetosphere of B \citep{bkk+08}. The spin precession rate predicted by general relativity means that it takes $75$~yrs for B's spin axis to complete a full precession cycle about the orbital angular momentum. This high precession rate is expected to have dramatic effects on B's pulse properties as a result of the change in apparent geometry with respect to our line-of-sight. These include the evolution of its pulse profile, light curves, and flux densities  \citep{bpm+05}.

The orbital plane of this system is almost edge-on to us. The orbital inclination was calculated to be $88.7\degr$ through timing \citep{ksm+06} and $89.7$$\degr$ through modeling of the interstellar scintillation properties of the two pulsars \citep{cmr+05}. Eclipses of A by B are a fortunate consequence of this favorable geometry.
The $2.9$ lt-s diameter of the orbit, determined through radio timing, and the 30-s duration of the eclipse, show that the magnetosphere of B is roughly 10\% of its light-cylinder radius \citep{lbk+04}.
The pulse profile of the more energetic pulsar A is stable. The limit on variations of A's profile width at 10\% of the pulse peak is only $0.5$$\degr$~yr$^{-1}$ \citep{mkp+05}.
This can be explained by a range of geometrical parameters. \citet{mkp+05} found that both circular and elliptical beams could fit the data, with a slight preference for an elliptical beam.
\citet{fsk+08} found that the misalignment between spin and orbital angular momentum axes of pulsar A is very low, with a $95$\% upper limit of $14$$\degr$, assuming emission from both magnetic poles. This supports the idea that the formation process of pulsar B was relatively symmetric.

The high spin-down energy loss rate of A likely causes the magnetosphere of B to be deformed in a way similar to that of the Earth by the Solar wind \citep{lyu04}.
The long tail will point towards the Earth around the eclipse of A by B, near the inferior conjunction of B (i.e. at an orbital phase of $270$$\degr$, as measured from the ascending node). We find that the bright emission from B is detected in two different orbital phase windows: at $\sim$$180$$\degr$$-$$240$$\degr$ (hereafter BP1), and $\sim$$260$$\degr$$-$$310$$\degr$ (hereafter BP2).
In addition to these two bright phases, the pulsar often shows weak emission at other orbital phases \citep{lbk+04}.
The exact explanation of this exceptional behavior is challenging. The model of \citet{lyu05} explains the phenomenon generally, but fails to produce the second bright phase at the correct orbital phase ($\sim$$245$$\degr$$-$$270$$\degr$ instead of $\sim$$260$$\degr$$-$$310$$\degr$). It also can not explain the evolution of the two bright phases (see \S 2.3).

Another effect of A's electromagnetic radiation on B's magnetosphere is drifting subpulses in the emission of B which appear similar to those seen for some normal radio pulsars \citep{mkl+04}. The drifting feature can be seen only in BP1 and not in BP2. The separation of the drift bands is equal to the period of A, showing that the drifting features are a direct result of the impact of A's $44$ Hz electromagnetic radiation on  B.
A possible explanation for the occurrence of drifting features only in BP1 is the particular geometric configuration of the two pulsars with respect to our line-of-sight. \cite{fwk+09} proposed a new technique for timing the double pulsar system by describing a geometrical model which predicts the delay of B's subpulses relative to A's radio pulses.

In this paper, we present a study of the extreme variation of B's pulse profile over time, which concludes with its total radio disappearance in March 2008. We present an analysis based on more than five years of data taken with the Green Bank Telescope. In $\S$\ref{obs} we present our observational data and our analysis methods. We also discuss the disappearance of B's emission and the evolution of the emission from the beam. In $\S$\ref{geomodel}, we estimate the geometrical parameters of pulsar B by using the empirical circular beam model of \citet{cw08} for the two-dimensional pulse profile of a precessing pulsar. Then we extend this model to an elliptical beam that explains the profile evolution well and results in realistic geometrical parameters.
In $\S$\ref{dis}, we discusses how the different types of beam shapes relate to the observed pulsar data and show that this is
the best explanation for the behavior of B.

\section{Observations and Analysis}
\label{obs}

We observed J0737$-$3039B with the $100$-m Green Bank Telescope (GBT) in West Virginia from $2003$ December $24$ to $2009$ June $20$. Our
observations are at multiple frequencies - because 820 MHz is the most common, we use only those data for this paper.
Until $2009$ January, the data were taken using the GBT spectrometer SPIGOT card with sampling time of $81.92$ $\mu$s.
 SPIGOT provides $1024$ synthesized frequency channels across a $50$ MHz bandwidth. After $2009$ January, the GBT spectrometer GUPPI was used. It covers $2048$ frequency channels with sampling time of $61.44$ $\mu$s, with a wider bandwidth of $200$ MHz. All the data were dedispersed and folded using the pulsar analysis package SIGPROC, assuming a dispersion measure of 48.914~cm$^{-3}$pc \citep{lbk+04}. We used the ephemeris of \citet{lbk+04} until 2006, and since  then have used the ephemeris of \citet{ksm+06} to form mean pulse profiles.
A total of $52$ data sets with $234$ total hours of observations at 820 MHz have been analyzed.

\subsection{Pulse profile evolution of the two bright phases}

\begin{figure*}
\epsscale{1.60}
\plotone{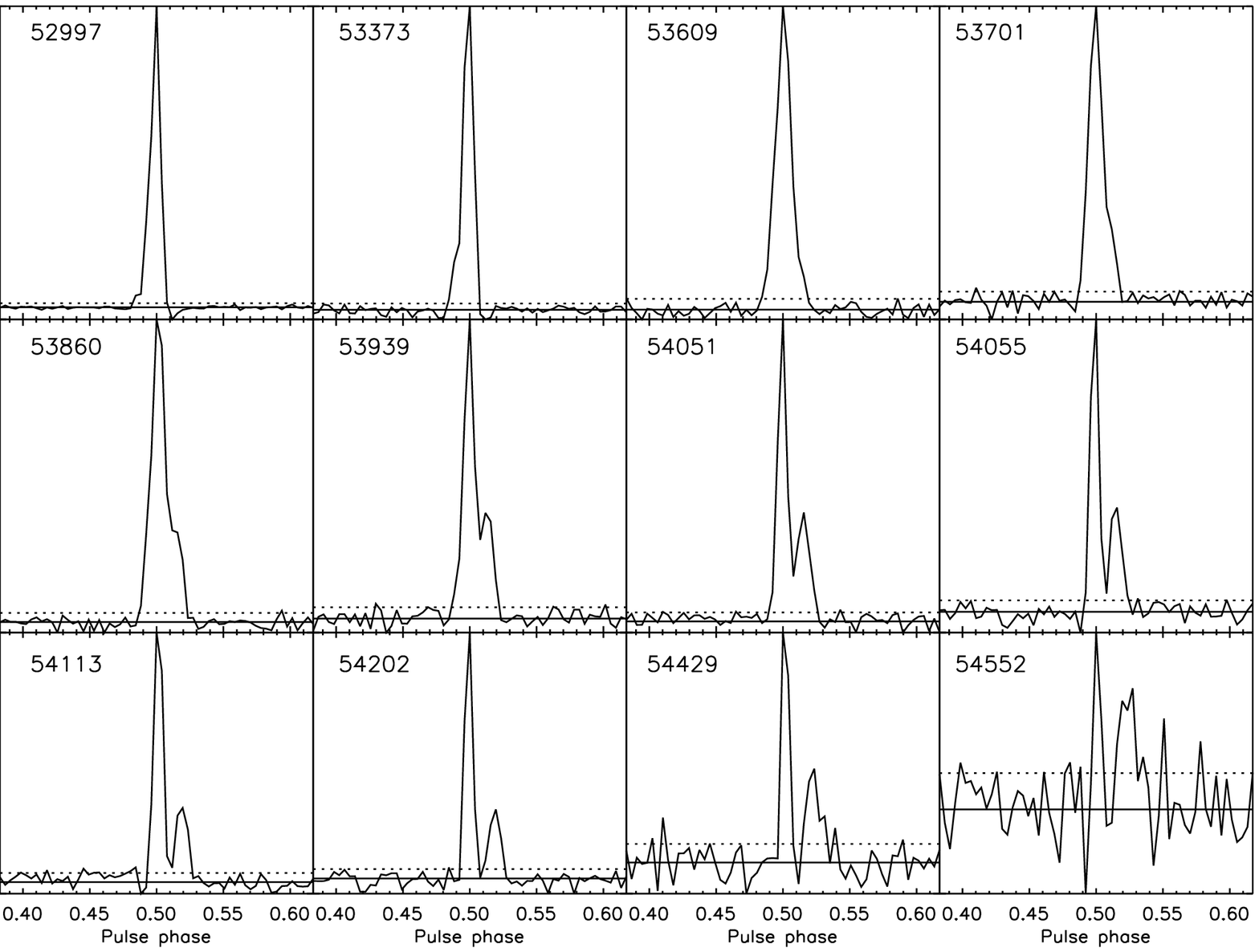}
\caption{
The pulse profiles of BP1 on 12 different days including the very first observation (MJD $52997$). All data have been observed at a frequency of $820$ MHz. Each profile covers 20 min of orbital longitude (from 185$\degr$$-$235$\degr$) and there are 256 bins across the entire profile. Since predictions of absolute pulse phase are not available for these observations, the highest profile peak has been aligned to phase 0.5 at each epoch. The horizontal solid and dotted lines show the baseline, or off-pulse mean, of the profile and the corresponding standard deviation of the off-peak region, respectively. The effective time resolution of the profiles is $0.01$ sec. 
\label{profile_phase1}}
\end{figure*}

\begin{figure*}
\epsscale{1.60}
\plotone{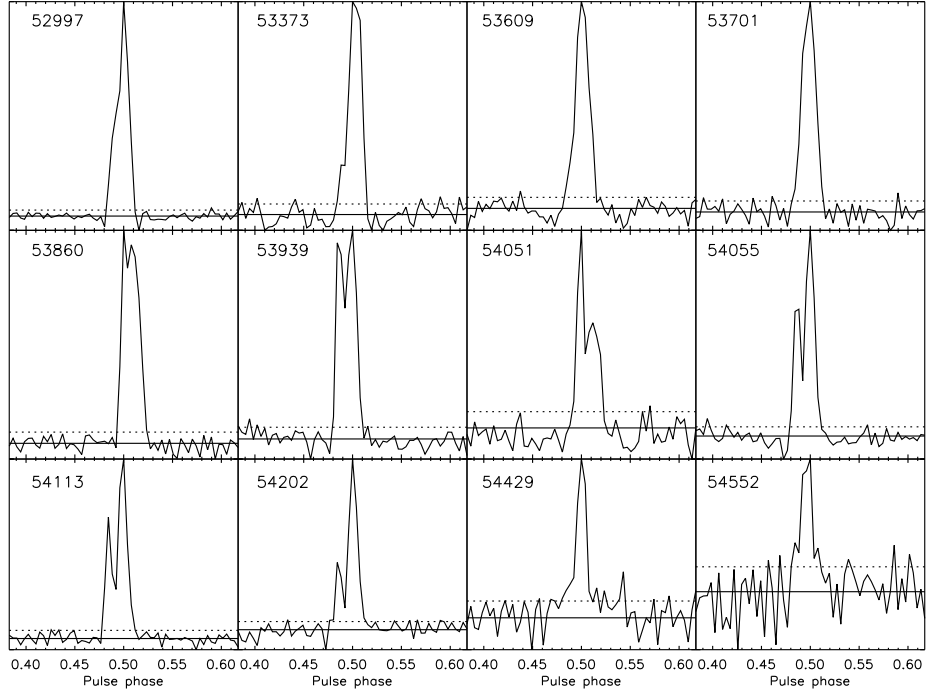}
\caption{ As described in Fig.~\ref{profile_phase1}, but for BP2.
\label{profile_phase2}}
\end{figure*}

The pulsar is brightest in the orbital phase region $185$$\degr$$-$$235$$\degr$, and second brightest in $265$$\degr$$-$$305$$\degr$ \citep{lbk+04,bpm+05}. Table~1 shows the longitudes (as measured from the ascending node) of each phase in which B is easily detected in early observations. In Fig.~\ref{profile_phase1} and \ref{profile_phase2}, we present the integrated pulse profiles for BP1 and BP2, respectively, on $12$ days that pulsar B was observed at $820$ MHz.  The very first observation, which was made in December $2003$, shows a unimodal average pulse profile.
This unimodal pulse profile gets broader and splits gradually into a two-component pulse profile over time. This evolution is common for both bright phases, but the specific features are somewhat different.

\begin{table*}
\caption{
Starting and ending orbital longitudes of each phase in which B is easily detected.}
\begin{center}
\begin{tabular}{lllll}
\hline
\multicolumn{1}{c}{} &
\multicolumn{1}{c}{BP1} &
\multicolumn{1}{c}{BP2} &
\multicolumn{1}{c}{WP1} &
\multicolumn{1}{c}{WP2} \\
\hline
Start & 185$\degr$ & 265$\degr$ & 340$\degr$ & 80$\degr$ \\
End & 235$\degr$ & 305$\degr$ & 30$\degr$ & 130$\degr$ \\
\hline \end{tabular}
\tablecomments{The two bright phases are denoted as BP1 and BP2 and the two weak phases as WP1 and WP2.}
\end{center}
\label{phases}
\end{table*}

\begin{figure*}
\epsscale{1.60}
\plotone{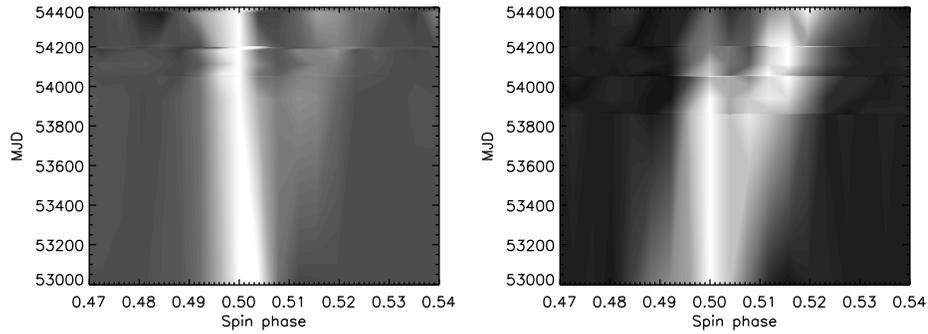}
\caption{Grey$-$scale plots of BP1 (left) and BP2 (right). There are 256 bins across each pulse profile. The first peak of each pulse profile has been normalized to unity and aligned to visualize how the second peak evolves. The discontinuities of the plots are artifacts due to boundary level changes.
\label{grey_scale}}
\end{figure*}

The evolution from a single-peak profile towards a two-peak profile begins around $2006$ May (MJD $53860$). In BP1, the first peak is brighter than the second one. This two-peak profile exists almost until the disappearance of the pulsar in this orbital phase region (see Fig.~\ref{profile_phase1}).
In BP2, the evolution has a similar pattern to BP1, but the remarkable difference is that the relative heights of the two peaks differ (see Fig.~\ref{profile_phase2}). It initially looks similar to the profile of BP1, but on MJD $54055$ the relative amplitudes of the two peaks change and the second one becomes more prominent. This persists throughout the evolution and we believe that this feature is due to the impact of A's emission on B, the effect of which changes over the period of the observation.
Grey$-$scale plots of the first and the second bright phases of the two-peak profile are shown in Fig.~\ref{grey_scale}. The first peak of each epoch has been normalized and aligned in both phases. It can be clearly seen that the two peaks are moving apart and the separation grows with time in both phases simultaneously.

\begin{figure*}
\epsscale{1.60}
\plotone{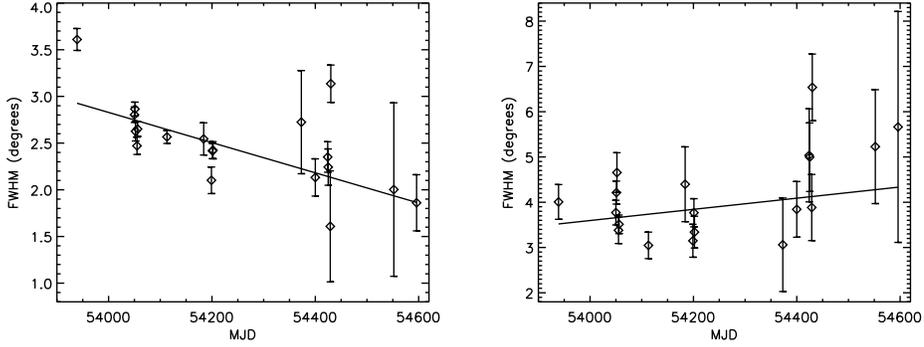}
\caption{
Full width at half max (FWHM) of the first peak of BP1 (left) and the second peak of BP1 (right).
The linear least-squares fits show a decreasing FWHM rate of $0.59(8)\degr$~yr$^{-1}$ for the first peak and increasing FWHM rate of $0.4(3)\degr$~yr$^{-1}$ for the second peak.
\label{fwhm_phase1}}
\end{figure*}

\begin{figure*}
\epsscale{1.60}
\plotone{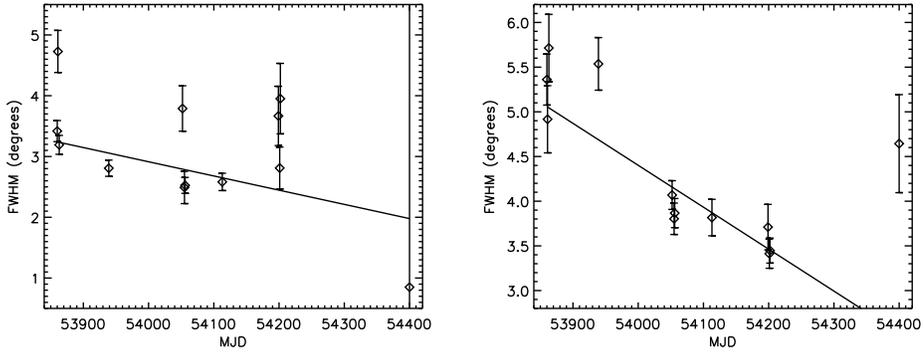}
\caption{
FWHM of the first peak of BP2 (left) and that of second peak of BP2 (right).
The linear least-squares fits show a decreasing FWHM rate of $0.8(2)\degr$~yr$^{-1}$ and $1.7(2)\degr$~yr$^{-1}$ for the first and second peak, respectively. Note that due to the small number of bright pulses included in these composite profiles, the profiles show a high amount of variability that does not follow any trend.
\label{fwhm_phase2}}
\end{figure*}

In order to get more information on the two peaks to describe the evolution, we fit two Gaussians separately for each peak of the two-peak pulse profile. From the Gaussian fits, we calculate the full-width-half-max (FWHM) of each peak of the two-peak pulse profiles in the two bright phases. They are shown in Figs.~\ref{fwhm_phase1} and \ref{fwhm_phase2}. Linear least-squares fits  qualitatively show that the FWHM of the first peak of BP1 decreases with time while that of the second peak increases slightly. In BP2, the FWHM of the first peak shows a slight decrease, but the decrease of second is more prominent.

\begin{figure*}
\epsscale{1.60}
\plotone{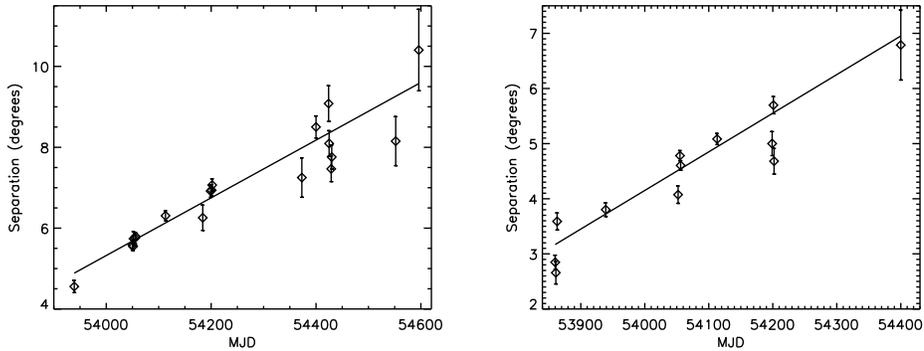}
\caption{
The varying separation of the two peaks of pulse profiles of BP1 (left) and BP2 (right). The linear least-squares fits show a separation rate of $2.6(1)$$\degr$~yr$^{-1}$ and $2.6(2)$$\degr$~yr$^{-1}$ for BP1 and BP2, respectively. Note that in BP1, the two-peak pulse profile lasts longer than in BP2.
\label{separation}}
\end{figure*}

 We present the evolution of the separation of the peaks in  the two bright phases in Fig.~\ref{separation}.The rates of separation are calculated from a linear least-squares fit to be $2.6(1)$$\degr$~yr$^{-1}$ and $2.6(2)$$\degr$~yr$^{-1}$ for BP1 and BP2, respectively. It appears that the profiles in both bright regions present the same rate of change in their component separation.

\subsection{Flux evolution of the two bright phases}

\begin{figure*}
\epsscale{1.60}
\plotone{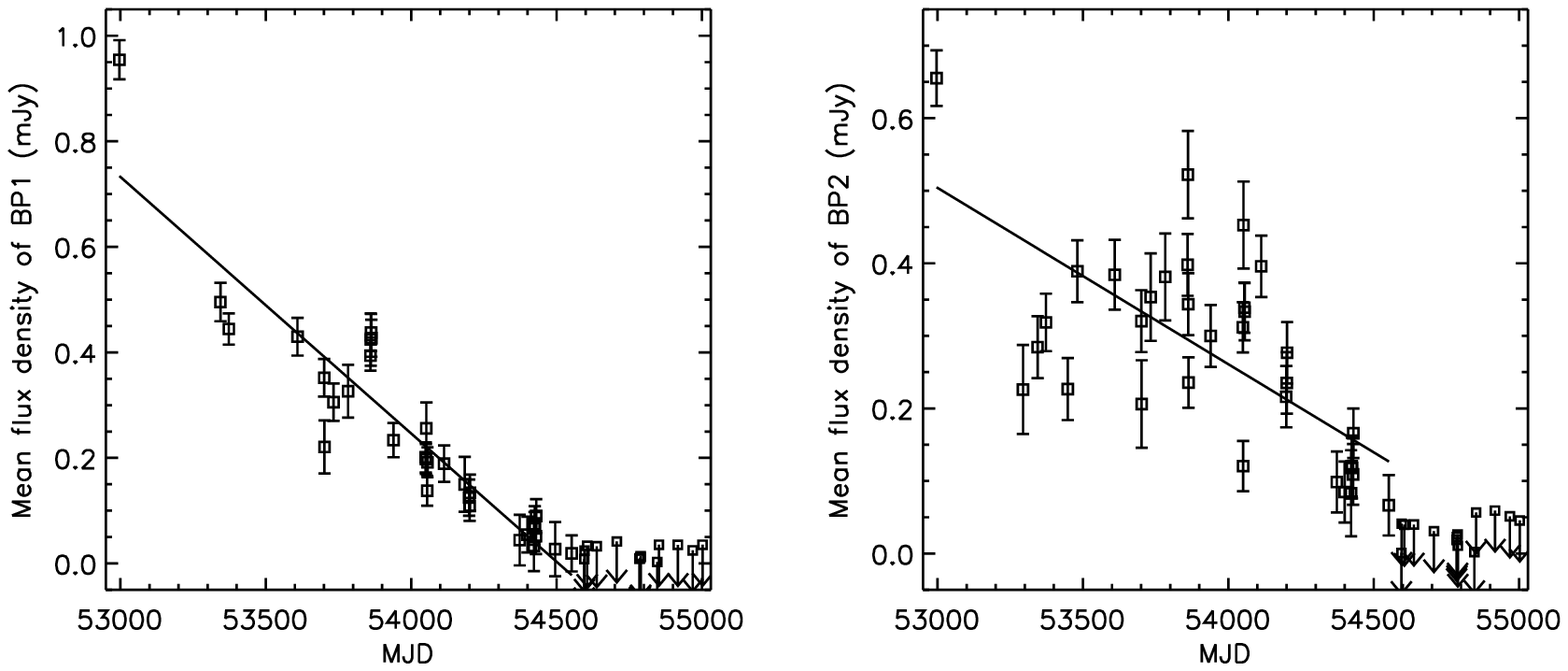}
\caption{
Mean flux density of the radio emission in BP1 (left) and in BP2 (right).
The data marked by the arrows are the upper limits and ignored in our linear least-squares fit.
The decreasing rate of the mean flux density is calculated to be 0.177(6) and 0.089(7) mJy~yr$^{-1}$ for BP1 and BP2, respectively.
\label{flux}}
\end{figure*}

In both bright regions, the integrated pulse flux density has decreased gradually over time (see Fig.~\ref{profile_phase1} and \ref{profile_phase2}). The pulsar was detected in both bright phases with the last significant detection in March 2008 (MJD 54552) at $820$ MHz.
We estimated flux densities at 820 MHz using the radiometer equation. First we calculated the radiometer noise, using a system temperature of 35~K (the system temperature is defined as the sum $T_{\rm sys} = T_{\rm rec} + T_{\rm spill} + T_{\rm atm} + T_{\rm sky}$), and considered this as the flux density at the off-pulse region
of the pulse profile. The flux density is then obtained by multiplying the pulse profile by the ratio of radiometer noise to the standard deviation of the off-pulse phase and subtracting the mean off-pulse level. We have carried out this calculation for only the two bright phases, because the emission in the weak phases disappeared much earlier (discussed in $\S2.4$ with more details). The calculated flux densities in both bright phases on MJD 52997 ($0.95(4)$ and $0.65(4)$ mJy for BP1 and BP2 at 820 MHz, respectively) are consistent with the value that has been calculated by \citet{lbk+04} (0$-$1.3(3) mJy at 1390 MHz). Fig.~\ref{flux} shows the mean flux densities of different epochs which have been observed at $820$ MHz. This confirms that the flux density significantly decreases over time and almost reaches zero around MJD $54852$ in both bright phases.
The rate of decrease is calculated to be $0.177(6)$ and $0.089(7)$ mJy~yr$^{-1}$ for BP1 and BP2, respectively. The flux densities of the last few epochs are only upper limits (denoted by arrows) and not included in the fits. Our timing solution is not stable enough to provide a reliable prediction of the expected phase of the pulsar on these days, making it difficult to determine whether any apparent peaks are real. The peaks on MJDs 54856 and 54852 have the same pulse phase in both bright phases, suggesting they are real. However, given their low signal-to-noise, we describe them by upper limits.

\subsection{Analysis and comparison of light curves of the two bright phases}

\begin{figure*}
\epsscale{1.60}
\plotone{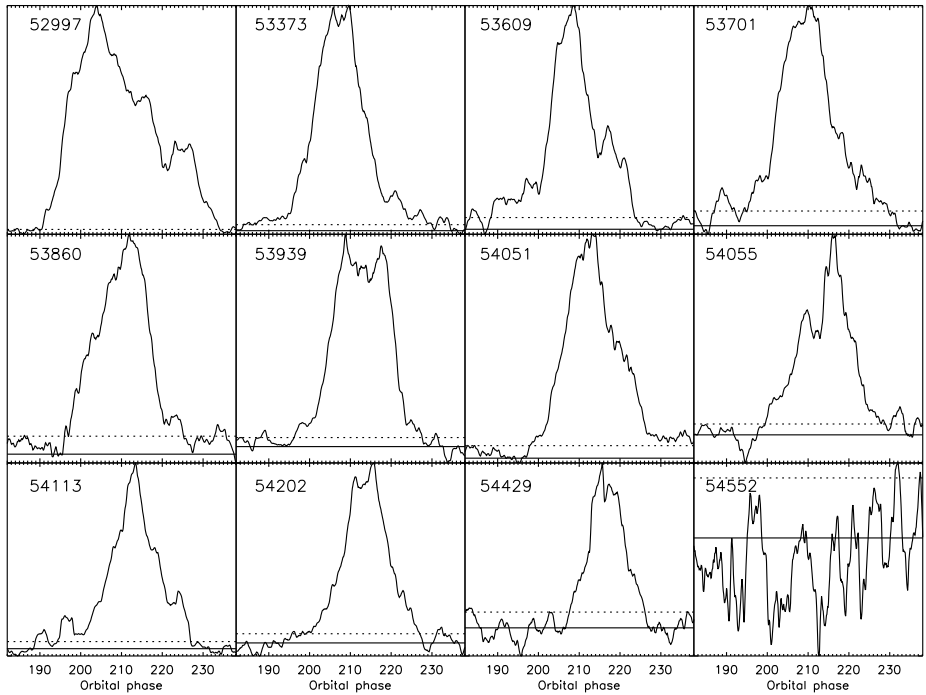}
\caption{
The light curves of BP1 on 12 different days at an observing frequency of $820$ MHz. Each data set has been cleaned to get rid of radio frequency interference. All light curves have been smoothed by using a boxcar averaging technique with a width of 30 pulses. The horizontal solid and dotted lines show the baseline of the plot and the corresponding standard deviation of the off-peak region (140$\degr$$-$180$\degr$ and 240$\degr$$-$260$\degr$), respectively. The last data set shows just noise.
\label{lcurve1}}
\end{figure*}

\begin{figure*}
\epsscale{1.60}
\plotone{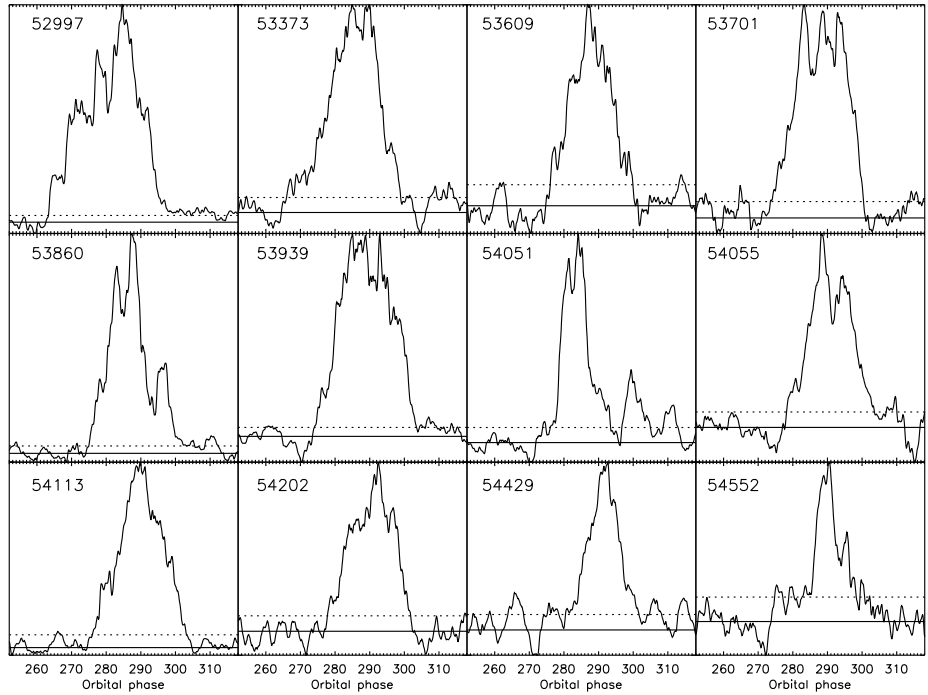}
\caption{ As in Fig.~\ref{lcurve1}, but for BP2.
\label{lcurve2}}
\end{figure*}

The orbital-phase binned light curves of the two bright phases were obtained by integrating the flux in a window covering 5\% of the spin period and centered on the pulse peak to reduce the effect of baseline noise. Then each light curve is smoothed by using a boxcar with a width of 30 pulses to reduce the significant pulse-to-pulse modulation.
In Figs.~\ref{lcurve1} and \ref{lcurve2}, we present the light curves of the two bright phases on $12$ different days that have been observed at frequency $820$ MHz. Initially in both bright phases the emission was detected in a region covering $\sim30$$\degr$ of orbital phase, but this has shrunk over time, although this did not happen symmetrically in both phases.

In order to analyze the evolution of the two bright phases with time, we calculated the start and end orbital longitudes, at the 10\% of the maximum, of BP1 and BP2. These are shown in Fig.~\ref{shrink}. The linear least-squares fits show the start of BP1 moves to higher longitudes at a rate of $3.1(4)$$\degr$~yr$^{-1}$ and the end of BP1 moves backwards at a rate of $-1.4(4)$$\degr$~yr$^{-1}$. In BP2, the both start and end orbital longitudes move to higher values at a rate of $4.1(2)$$\degr$~yr$^{-1}$ and $2.1(2)$$\degr$~yr$^{-1}$. These rates show that BP1 and BP2 shrink at a rate of $4.5(6)$$\degr$~yr$^{-1}$ and $2.1(3)$$\degr$~yr$^{-1}$, respectively. The above rates are updates for those presented by \citet{bpm+05}. These results confirmed that the disappearance did not happen symmetrically. However, the disappearance did occur simultaneously in both bright phase regions.

\subsection{Analysis of the two weak phases}

Weak radio emission has been detected in two orbital phase ranges in addition to the two bright phases. One of the two weak emission regions, $340$$\degr$$-$$30$$\degr$ (hereafter WP1) is much brighter than the other, $80$$\degr$$-$$130$$\degr$ (hereafter WP2), and is detectable until $2007$ November (MJD $54429$) (see Fig.~\ref{profile_weak1}). WP2 vanished in $2006$ February (MJD 53783), more than a year before the disappearance of the WP1 (see Fig.~\ref{profile_weak2}). In general, both weak phases show a unimodal pulse profile and no dramatic pulse shape changes over time.
 Some data show two apparent peaks in their pulse profile, but the lower signal-to-noise makes it difficult to model these in the same way as for the bright phases (e.g. on MJD $53702$ in WP1 and $53701$ in WP2). If the evolution is solely due to relativistic spin precession, we would  expect the same profile evolution for all phases. However, the emission may not be strong enough in the weak phases to detect the second profile component. For instance, at MJD 54113 the brightness of the second brightest component of BP1 is 30\% of the brightest component and that of BP2 is 67\%. In comparison, the root-mean-square noise in WP1 is almost 50\% of the pulse maximum, making it difficult to detect the second peak of the profile. Therefore, within the uncertainties, the component separation evolution may be similar for bright and weak phase profiles.

We calculated the mean flux densities of both weak phase regions (see Fig.~\ref{flux_WP}).
In these phases, the mean flux densities are lower than that of bright phases. The decreasing rates of the mean flux densities are calculated to be 0.032(8) and 0.02(2) mJy~yr$^{-1}$ for WP1 and WP2, respectively. The larger uncertainty of the data points results in the poor quality of the fits. However, due to the weakness of the pulse emission in WP1 and WP2, we only use the two bright phase regions for our modeling.

\section{Determining the Geometry of the system}
\label{geomodel}

\citet{cw08} proposed a framework to model the 2-D pulse profile of a precessing pulsar by using a simple circular hollow-cone beam. They applied this model for PSR B1913$+$16 to explain its long term pulse profile evolution. To represent the different intensity levels of the emission beam, they constructed a set of coaxial circular hollow cones with different angular radii. Each circular cone represents a different intensity level. The outermost cone represents the lowest intensity level. The intensities gradually increase towards the maximum-intensity-level cone cone somewhere between the outermost cone and the center of the beam, and then decrease until reaching the center of the beam. With this special construction of the beam, it is possible to explain the hour-glass or oval shapes of the 2-D pulse profiles.

Moreover, this model constrains the geometrical parameters of the precessing pulsar.
Mainly, the model predicts the longitudinal separation (i.e., pulse width) at a given epoch for a given intensity level for some set of geometrical parameters. Then we fit this model-predicted width to the corresponding observed-pulse-profile width for a range of intensity levels to obtain the geometrical parameters.

First we used the hollow-cone beam shape in order to determine the geometry of the B pulsar. Then we used a more complicated elliptical horse-shoe shaped beam, filled with different intensity levels as in the circular beam construction, to constrain the geometry and found that this fit our observations better.

In both cases, the required pulse-width data at different intensity levels of the pulse profiles have been calculated by fitting Gaussians to the profiles and then calculating the pulse widths at different intensity levels of the fitted Gaussian curves. The modeling has been done for the two bright phases separately to constrain two sets of geometrical parameters.

\subsection{Modeling the geometry with a circular hollow-cone beam}

In this section, we use a circular hollow-cone beam and the same set of equations as in CW08 to determine the two angles $\alpha$ and $\theta$, which are the magnetic inclination (i.e. angle between spin axis and magnetic dipole moment) and the colatitude of the spin axis (i.e. angle between spin axis and orbital angular momentum), respectively (see Fig.~\ref{fig_dig} (a)). The model also includes another parameter, T$_{0}$, the epoch for which the precession phase is equal to zero.
The fit of the width at different intensity levels of our data to the model is shown in Fig.~\ref{model}. The fit has been done by minimizing the chi-square statistic over the entire parameter space of the two angles ($\alpha$:[0,$\pi$], $\theta$:[0,$\pi$]) and T$_{0}$:[1930,2001]. This range of dates encompasses one full precessional cycle for pulsar B. The required errors of the data points for the chi-square statistic are obtained by the fitted Gaussians, which are the same for a given epoch at different intensity levels.

The best-fit results for the two angles for BP1 are $\alpha$ $=$ $80.0$$\degr$$_{-15.0^{\circ}}^{+1.5^{\circ}}$ and $\theta$ $=$ $20.0$$\degr$$_{-16.5^{\circ}}^{+1.5^{\circ}}$, while for BP2 $\alpha$ $=$ $78.0$$\degr$$_{-11.0^{\circ}}^{+1.5^{\circ}}$ and $\theta$ $=$ $20.0$$\degr$$_{-12.5^{\circ}}^{+1.0^{\circ}}$. A secondary peak in the PDFs caused large errors for the best fit parameters at their 68.5\% confidence interval. The fits estimate T$_{0}$ $=$ 1959.4(1)~yr for both phases. These best fit values are very different from those in \citet{bkk+08}, who determined  $\alpha$ $=$ $70.9(4)$$\degr$ and $\theta$ $=$ $130.0(4)$$\degr$ (see Table~2). In addition to this, the simple circular beam shape predicts a disappearance in $\sim$ 2018 instead of 2008. Due to the poor fit and the inability to explain the disappearance of the radio emission in 2008, we experimented with other possible beam shapes.

\subsection{Modeling the geometry with an elliptical horse-shoe shaped beam}

In this case,
we used an elliptical hollow-cone filled beam with different intensity levels and then assumed that only a part of this beam is detectable, resulting in a horse-shoe shaped beam.
To simplify the geometry, we assumed that the beam shape is fixed with respect to the neutron star and it rotates around the spin axis in a way such that the semi-major axis is always aligned in the radially outward direction from the spin axis (see Fig.~\ref{fig_dig_new}).
In order to model this particular shape, we derived an equation for the profile width

\begin{equation}
w_{j}(t) = \arcsin[2\sqrt{[n_{x}(t)]^{2}+[n_{y}(t)]^{2}}\sin\eta_{j}(t)],
\end{equation}

\noindent
where {\it w$_{j}$(t)} is the width of the pulse profile at a given time, with the subscript {\it j} specifying the intensity level of the profile. The line-of-sight vector, {\it $\hat{n}(t)$}, can be expressed as in equation (3) of CW08 and {\it $n_{x}(t)$} and {\it $n_{y}(t)$} in the above equation are the X and Y components of {\it $\hat{n}(t)$}. The third variable, {\it $\eta_{j}$(t)}, is the angle which is subtended at the origin of the coordinate system due to the encounter of the line-of-sight vector with the elliptical beam; namely, it is the angle between the line joining the origin with the point at which the line-of-sight vector encounters the beam and the line joining the origin with the center of the elliptical beam (see Fig.~\ref{fig_dig_new}). Thus, amending the expression of CW08 for an elliptical beam, {\it $\eta_{j}$(t)} can be expressed as

\begin{equation}
\eta_{j}(t) =
{\rm acos}\big[\frac{n^{2}(t)+\sin^{2}\alpha - [n(t)-\sin\alpha]^{2}[1-R_j^{2}] + b^{2}_{j}}{2n(t)\sin\alpha}\big]
\end{equation}

\noindent
where $R_j =   b_{j}/a_{j}$ and $a_{j}$ and $b_{j}$ are the semi-major and semi-minor axes of the elliptical beam for a given intensity level. They can take different values corresponding to the intensity levels of the beam in a way such that  $a_{j}/b_{j}$, or the ellipticity of the beam, is a constant.

The horse-shoe shaped beam is a subsection of the elliptical beam. We can construct this by restricting the detectable area of the beam to a section of the elliptical beam, with {\it r$_{1}$} and {\it r$_{2}$} the radial distances on the X-Y plane of the elliptical beam within which emission is detectable (see Fig.~\ref{fig_dig} (b)). Simply, we can detect the emission beam of the pulsar only when $\sin\alpha + r_{1} < \sqrt{[n_{x}(t)]^{2}+[n_{y}(t)]^{2}} < \sin\alpha + r_{2}$, where $n_{x}(t)$ and $n_{y}(t)$ are the X and Y components of $\hat{n}(t)$.

The fit has been done for the two bright phases separately by searching over the entire parameter space for the two angles $\alpha$ and $\theta$ and the ratio $a/b$ using a likelihood analysis (see, e.g., \citet{mc00}). First the individual likelihood of each data point was calculated by assuming a Gaussian distribution such that the likelihood for model $\Theta$ and measurement $i$ is ${\cal L}_i(\Theta) = (2\pi\sigma_i^2)^{-1/2} {\rm exp}(-(w_{{\rm model},i}-w_{{\rm measured},i})^2/2\sigma_i^2)$, where $w_{{\rm model},i}$ and $w_{{\rm measured},i}$ are the model and measured widths and $\sigma_i$ is the error on the measured width. We then multiplied all ${\cal L}_i(\Theta)$ for all data points $i$ (i.e. profiles from all epochs and at all intensity levels) to get the total likelihood ${\cal L}(\Theta)$ for parameter combination $\Theta$. ${\cal L}_{\rm tot}$, the total likelihood over all parameter combinations, was calculated by summing all the likelihoods over all parameter combinations. Then the posterior probability distributions for parameter values were calculated by dividing the likelihoods for a particular parameter value by the total likelihood ${\cal L}_{\rm tot}$, assuming flat prior probability distributions. The best fit 2-D pulse profile is shown in Fig.~\ref{mymodel} and the posterior probabilities of the three parameters $\alpha$, $\theta$, and a/b for BP1 are shown in Fig.~\ref{like_error}.

The fitting constrained the two angles for BP1 to be $\alpha$ $=$ 65.0$\degr$$_{-1.0^{\circ}}^{+2.0^{\circ}}$ and $\theta$ $=$ 134.5$\degr$$_{-6.0^{\circ}}^{+1.5^{\circ}}$, while for BP2 $\alpha$ $=$ 73.0$\degr$$_{-3.5^{\circ}}^{+1.4^{\circ}}$ and $\theta$ $=$ 138.5$\degr$$_{-7.5^{\circ}}^{+4.0^{\circ}}$. We expect the same estimates for the two angles in both bright phases, indicating that our errors are underestimated or the model may not fit the data perfectly. These angles are similar to those derived by \citet{bkk+08} (see Table~2). The best-fit estimate of the ratio of the beam semi-major and semi-minor axes ($a/b$) is 1.72$_{-0.20}^{+0.12}$ and 1.15$_{-0.22}^{+0.27}$ for BP1 and BP2, respectively. Because of the varying influence of A,  the magnetosphere could have different shapes at different orbital phases, and these angles could in fact differ. In order to cause the disappearance in 2008, the ratio $r_{1}/r_{2}$ has to be 0.70 and 0.48 for BP1 and BP2, respectively. The parameter T$_{0}$ is 1950.3(1)~yr for both bright phases. Since the spin axis takes 75~yr to precess a full cycle around the orbital angular momentum axis, the year $2008$, within which disappearance occurred, is not a special year relative to this estimated T$_{0}$.

\section{Discussion}
\label{dis}

Pulse profile changes have  been detected in several other double neutron star systems: PSR B1913$+$16 \citep{wrt89,kra98,wt05}, B1534+12 \citep{arz96,sta04}, J1141$-$6545 \citep{hbo05,mks+10}, and J1906+0746 \citep{lsf+06}.
In  these pulsars, the long-term pulse profile evolution has been interpreted as the variation of the radio beam orientation with respect to our line-of-sight due to geodetic precession. \citet{kra98} demonstrated a fully consistent model for PSR B1913$+$16 assuming a circular hollow-cone beam geometry and the precession rate predicted by general relativity. More recently, CW08 revisited the problem with a different geometrical framework using the circular beam approximation.

We used the CW08 model in order to constrain the geometry of pulsar B.
First we used the circular hollow-cone beam shape they proposed in their paper and estimated the geometrical parameters $\alpha$ and $\theta$ by fitting our data.
The best fit values are different from the measurements of \citet{bkk+08} and the fit is poor.
It is likely that the discrepancy between our best-fit geometry and that of \citet{bkk+08}, as well as the inability of the CW08 to fully explain our data, are due to a non-circular emission beam shape for pulsar B.

\begin{table*}
\caption{
Geometrical parameters of pulsar B derived from our circular beam model (CBM) and elliptical beam model (EBM) and the eclipse model fitting by \citet{bkk+08}.}
\label{models}
\begin{center}
\begin{tabular}{lll}
\hline
\multicolumn{1}{c}{} &
\multicolumn{1}{c}{$\alpha(\degr)$} &
\multicolumn{1}{c}{$\theta(\degr)$} \\
\hline
CBM     :BP1 & 80.0($-15.0$, $+1.5$) & 20.0($-16.5$,$+1.5$)\\
CMB     :BP2 & 78.0($-11.0$, $+1.5$) & 20.0($-12.5$,$+1.0$)\\
EBM     :BP1 & 65.0($-1.0$, $+2.0$) & 134.5($-6.0$,$+1.5$)\\
EBM     :BP2 & 73.0($-3.5$, $+1.4$) & 138.5($-7.5$,$+4.0$)\\
\citet{bkk+08} & 70.9($-0.4$, $+0.4$) & 130.0($-0.4$,$+0.4$)\\
\hline \end{tabular}
\end{center}\end{table*}

In $\S$3.2, we modified the geometrical framework of CW08 with an elliptical hollow-cone beam shape. This produced a much better fit than the circular beam. Furthermore, the two angles $\alpha$ and $\theta$ are consistent in the two bright phases and with the predictions of \citet{bkk+08} (see Table~2).
 To obtain the disappearance in 2008, the detectable area of the beam has to be restricted to a section of the elliptical beam, producing a horse-shoe shaped beam.
Given this model, we calculate the  ratio of the semi-major and semi-minor axes ($a/b$) to be 1.72$_{-0.20}^{+0.12}$ and 1.15$_{-0.22}^{+0.27}$ for BP1 and BP2, respectively. After transforming this ratio to the assumed unit sphere by dividing by $\cos$($\alpha$), we estimate the ellipticity of the radio beam of B in the two bright phases separately: 0.9(1) and 0.9(2) at BP1 and BP2, respectively. The ratio $r_{1}/r_{2}$ in the two bright phases reveals that only 29(6)\% of the beam area was detectable in BP1 and only 21(5)\% in BP2.
The different estimates for the effective area of the beam in the two bright phases show that the emission beam may have different orientations corresponding to the different orbital phases. The characteristic pulse shapes are also different at different orbital phases. This is a direct consequence of the influence of pulsar A's wind on the emission beam of pulsar B (see, e.g., \citet{lyu04}).

The special choice of the beam construction with different intensity levels explains the single-peaked to double-peaked pulse profile evolution. On the first day, MJD 52997, the observed bright single-peaked pulse profile can be explained as our line-of-sight just grazing the outer edge of the maximum intensity level of the beam. When the line-of-sight gradually moves inward with time (i.e., towards the spin axis; see Fig~\ref{fig_dig}(b)), the two-peaked pulse profile is formed since the maximum intensity level of the beam crosses our line-of-sight twice. However, the observed decreasing flux density over time cannot be explained by either model, since the intensity is constant along the cross-section of a hollow cone of the conical beam, in both the circular or elliptical cases. We believe the observed decrease in flux density must be due to a gradient in brightness across the beam or due to the changing influence of A on B during the span of these data. The distance between the two pulsars changes over time relative to our line-of-sight due to the relativistic advance of periastron of 17$\degr$~yr$^{-1}$ of the system. This results in a different amount of impact on B's emission by A with respect to Earth and leads to a change in the shape of the horse-shoe beam slightly, but this is likely a small effect, and not easy to model. Thus we ignored this effect.

The data show that the pulse profile had two peaks when the disappearance occurred in 2008 and clearly did not become a single peak before the disappearance. The horse-shoe shaped beam model can explain this phenomenon qualitatively, as the line-of-sight moved out of the elongated side of the beam (i.e. the two opened arms of the horse-shoe beam, see Fig.~\ref{fig_dig} (b)). This results in the disappearance of both peaks of the pulse profiles in both bright phases simultaneously. The reappearance of the emission of pulsar B is predicted to occur in $\sim$ 2035, according to the horse-shoe beam model, assuming emission from the same part of the beam. This is because our line-of-sight encounters the emission beam twice during the time that the spin axis rotates around the orbital angular momentum axis once ($\sim$71~yr). This can be understood by visualizing the geometry of the emission beam with two angles, $\alpha$ $\approx$ $65$$\degr$ and $\theta$ $\approx$ $130$$\degr$. Thus the spin axis of the pulsar B is below the inclination plane and the magnetic moment axis crosses the inclination plane twice during every 2.8 sec period.
Since the inclination plane of the system is almost edge-on with respect to the Earth, we are able to detect the emission beam twice during the 71~yr period. Geodetic precession along with the unique  geometry  of the system is responsible for this phenomenon.
Note that this prediction for the reappearance time is only valid assuming a single horse-shoe emission component. If the beam is symmetric, with two horse-shoe shaped emission components separated by a gap, then we would expect a reappearance around $2014$.
Because we have only sampled a limited part of the beam, it is impossible to know what the true intensity distribution of B's beam is at this point.

Despite the unusual modulation of B's emission by A, the 3-5\% duty cycle of B's pulse and the pulse profile shapes throughout the evolution are similar to those of pulsars with similar periods, with the dramatic geodetic precession of B offering us the chance to sample a much larger portion of the beam than typically possible. It is difficult to determine what is responsible for B's flux density variation. It may be that most pulsars have instrinisic intensity variations across their beams.
 Or, the decrease in the intensity of B's emission with time may be solely due to B's magnetosphere shape changing with the varying influence of A. Theoretical modeling of the influence of A's wind with time is necessary to address this.
Our horse-shoe shaped beam fits the standard conal emission model \citet{ran83}, with the observed single to double peak evolution due to sampling different lines-of-sight across the conical beam, but this model does not explain the decreasing flux density and eventual disappearance of the pulsar. Pulse profile shapes have also been explained through the `patchy' beam model \citep{lm88} in which intensities across the beam vary widely and often only a portion of the cone is visible.
Observations of relativistic binary pulsar system J1141$-$6545 \citep{mks+10} show  that the emission across the beam is asymmetric with respect to the magnetic axis, consistent with a patchy beam.
It is more difficult to interpret the evolution of B's profile with the patchy beam model, however, unless the patches are large enough to explain both components disappearing at the same time. Continued monitoring and sensitive searches for emission from B will provide invaluable information on the true beam shape of this pulsar.  

\begin{figure*}
\epsscale{1.40}
\plotone{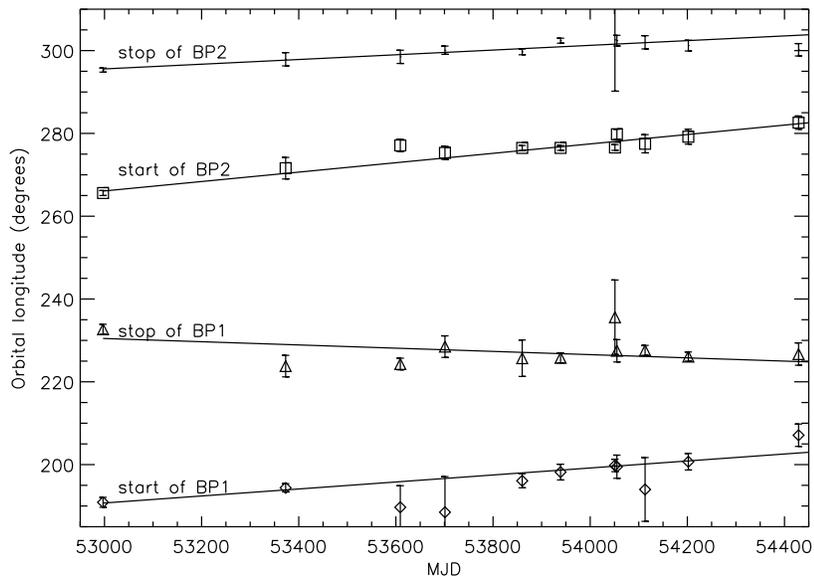}
\caption{Evolution  of the start and end orbital longitudes of BP1 and BP2, measured at the 10\% of the maximum.
\label{shrink}}
\end{figure*}

\begin{figure*}
\epsscale{1.60}
\plotone{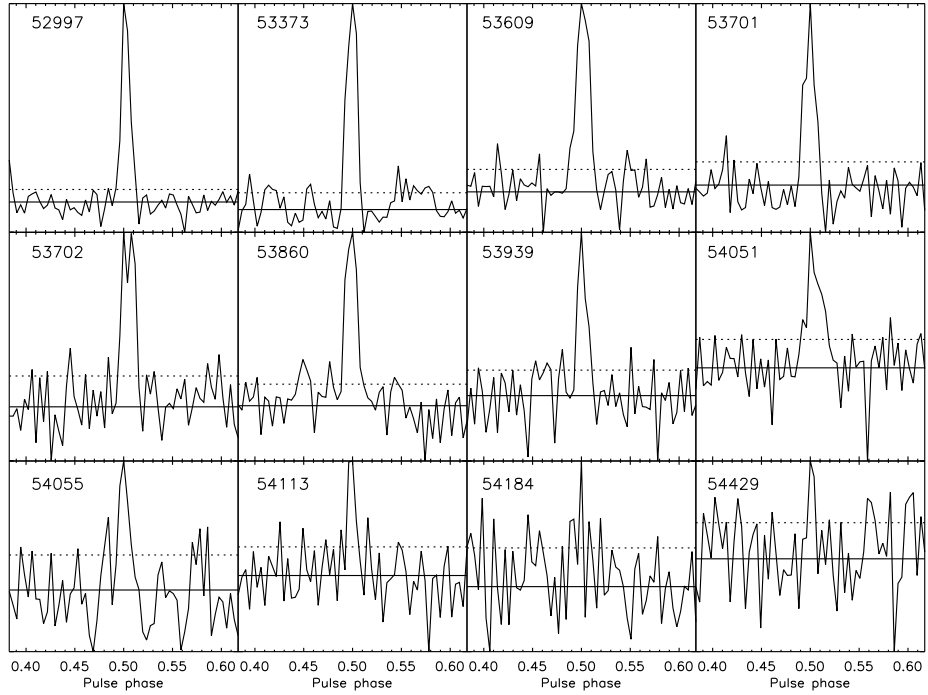}
\caption{
As in Fig.~\ref{profile_phase1}, but for WP1.
Each profile covers 20 min of orbital longitude (from 340$\degr$$-$30$\degr$) and there are 256 bins across the entire profile.
\label{profile_weak1}}
\end{figure*}

\begin{figure*}
\epsscale{1.60}
\plotone{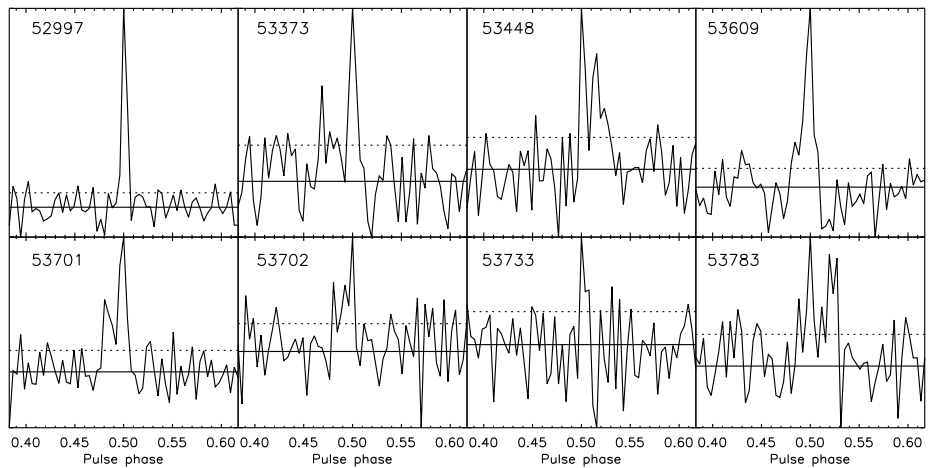}
\caption{ As in Fig.~\ref{profile_phase1}, but for WP2 covering 20 min of orbital longitude (from 80$\degr$$-$130$\degr$). The pulsar is weaker during this phase than during WP1.
\label{profile_weak2}}
\end{figure*}

\begin{figure*}
\epsscale{1.60}
\plotone{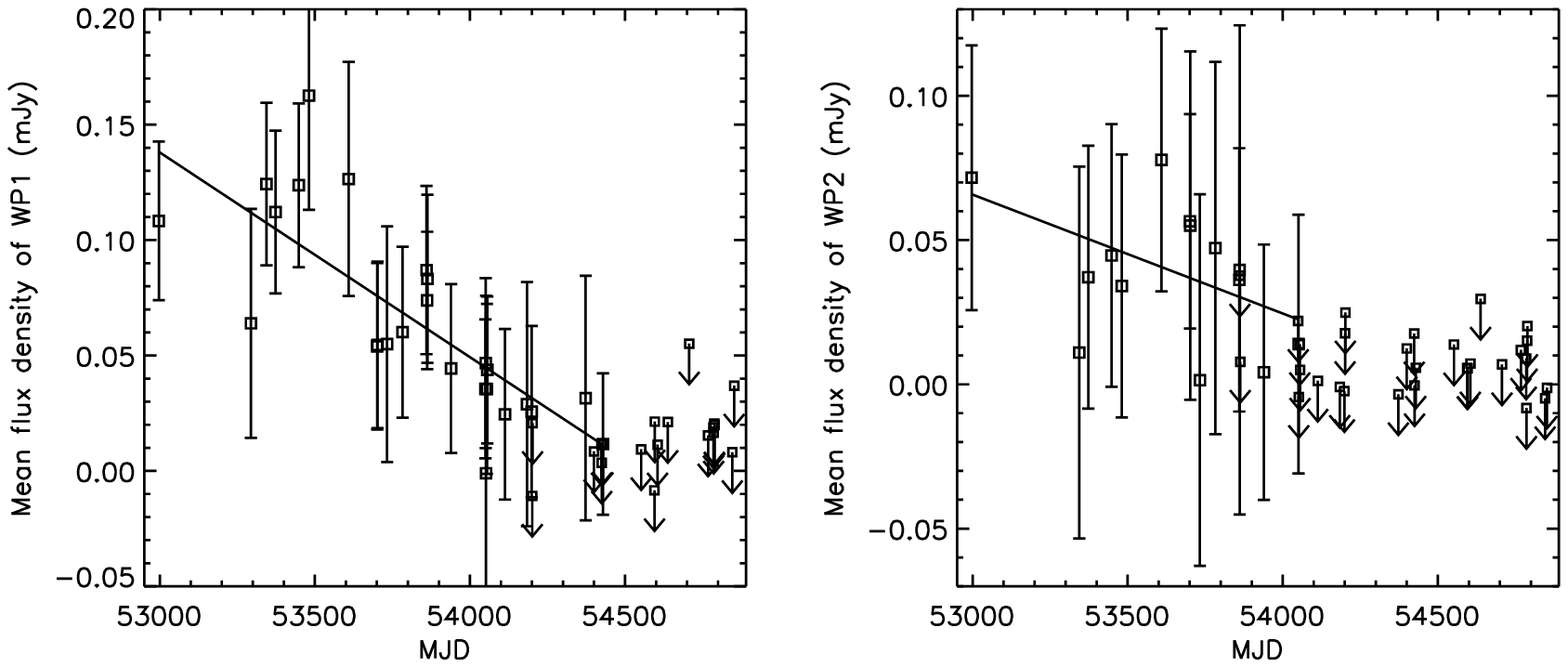}
\caption{
Mean flux density of the radio emission in WP1 (left) and in WP2 (right).
The data marked by the arrows are the upper limits and ignored in our linear least-squares fits.
The decreasing rate of the mean flux density is calculated to be 0.032(8) and 0.02(2) mJy~yr$^{-1}$ for WP1 and WP2, respectively.
\label{flux_WP}}
\end{figure*}

\begin{figure*} 
\epsscale{1.70}
\plotone{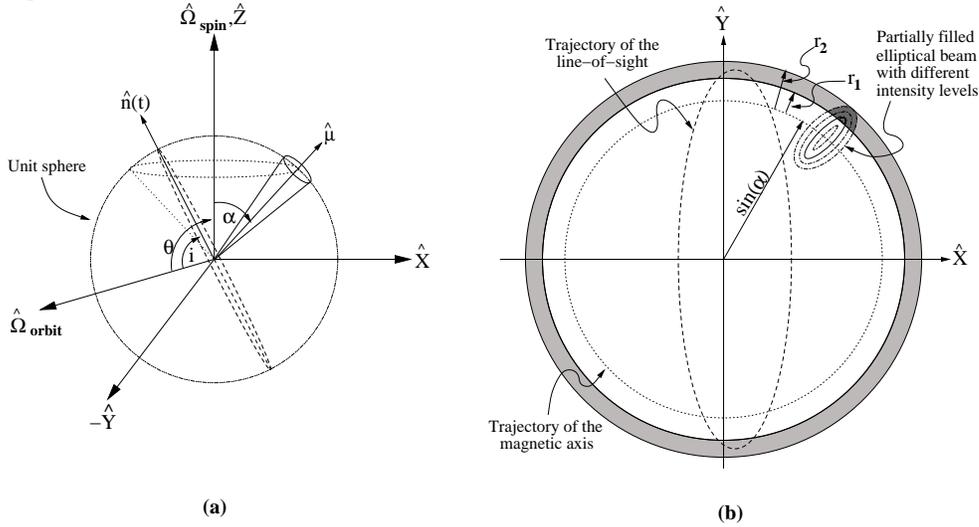}
\caption{Geometry of  pulsar B with its spin and the precession in a Cartesian coordinate system with a unit sphere at the center; all of the trajectories move on this sphere. {\bf (a)} The coordinate system is chosen with the Z-axis parallel to the spin-axis of the pulsar ($\hat{\Omega}$$_{spin}$) with the X-axis in the plane of the spin-axis and the orbital angular momentum axis ($\hat{\Omega}$$_{orbit}$); i.e., the $\hat{\Omega}$$_{spin}$ and the $\hat{\Omega}$$_{orbit}$ are always in the X-Z plane. The pulsar-earth line-of-sight is $\hat{n}(t)$ and dashes represent the conical trajectory of it on the unit sphere about $\hat{\Omega}$$_{orbit}$ due to geodetic precession. Dots represent the conical trajectory of the co-rotating magnetic field ($\hat{\mu}$) on the unit sphere about $\hat{\Omega}$$_{spin}$ due to the pulsar spin. The inclination of the orbit is {\it i}. The colatitude of spin axis and the misalignment of the magnetic field are {\it $\theta$} and {\it $\alpha$}, respectively. {\bf (b)} The projections of trajectories of $\hat{\mu}$ and $\hat{n}(t)$ on the X-Y plane. Dash-dot lines represent the different intensity levels of the elliptical beam. The intensity increases gradually from the outer most edge of the beam to inwards until the maximum intensity level (the solid line of the beam), and then decreases towards the center of the beam. The lightly shaded region represents the detectable beam area when the beam sweeps about the spin axis. This area is determined from the radial lengths $r_{1}$ and $r_{2}$, and they are measured from the center of the beam on the X-Y plane. With this restriction, the effective beam has a horse-shoe shape and it is represented as the dark shaded region. Dashes show the trajectory of the $\hat{n}(t)$ on the X-Y plane. When $\hat{n}(t)$ crosses the beam area (lightly shaded area), we are able to detect the pulsar.
\label{fig_dig}}
\end{figure*}

\begin{figure*}
\epsscale{1.60}
\plotone{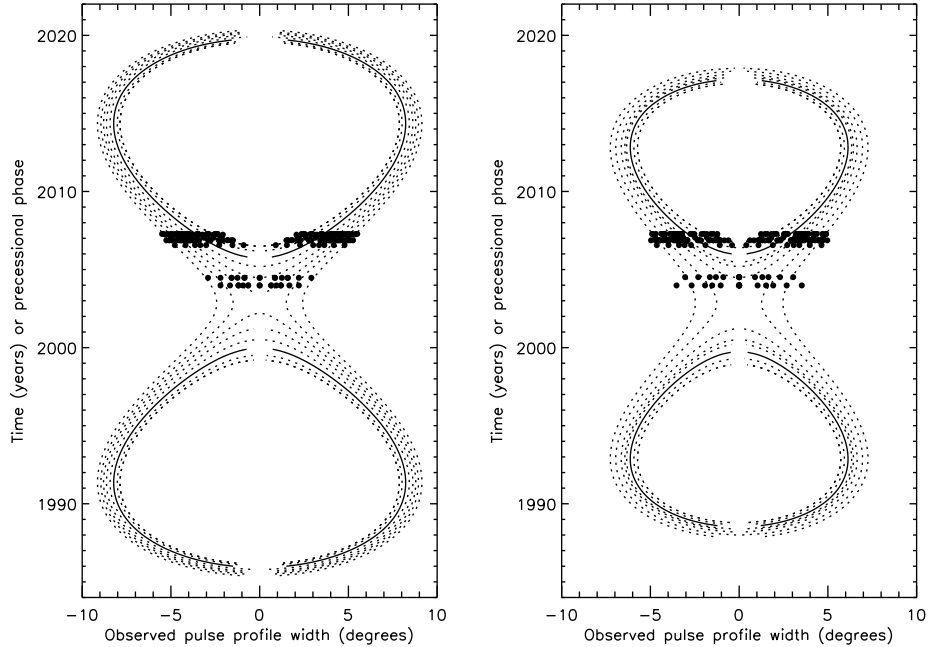}
\caption{
The hour-glass shaped two-dimensional pulse profile of PSR J0737$-$3039B, assuming a circular hollow-cone beam. The best fit parameters for  BP1 are $\alpha$ $=$ $80.0$$\degr$$_{-15.0^{\circ}}^{+1.5^{\circ}}$ and $\theta$ $=$ $20.0$$\degr$$_{-16.5^{\circ}}^{+1.5^{\circ}}$ (left), and for BP2 are $\alpha$ $=$ $78.0$$\degr$$_{-11.0^{\circ}}^{+1.5^{\circ}}$ and $\theta$ $=$ $20.0$$\degr$$_{-12.5^{\circ}}^{+1.0^{\circ}}$ (right). Fitting has been done by  searching the entire range of the two angles. The dots are the widths at equal intensity of the pulse profile. Each horizontal row of dots represents an observation at a given epoch. Equal-intensity contours are produced from the emission of the circular symmetric conical beam. The solid lines are the equal intensity level of the peaks of two-peak pulse profile. The intensity increases from the inner dashed line  outwards until the first solid line, which is the intensity of the peak, and then decreases outwards again. The intensity levels are, from inner dashed line, 80\%, 90\%, 100\%, 90\%, 80\%, 70\%, 60\% and 50\%. The vertical axis is calibrated in years, and can also be considered  the spin precession phase, which increases linearly with time. The time of precessional phase zero in the fits is T$_{0}$ $=$ 1959.41~yr for both phases. We have omitted the data between MJDs $53175$ and $53939$. During this time, the profile was bi-modal but the separation between the peaks was small, making it difficult to measure widths at different intensity levels.
\label{model}}
\end{figure*}

\begin{figure*}
\epsscale{.90}
\plotone{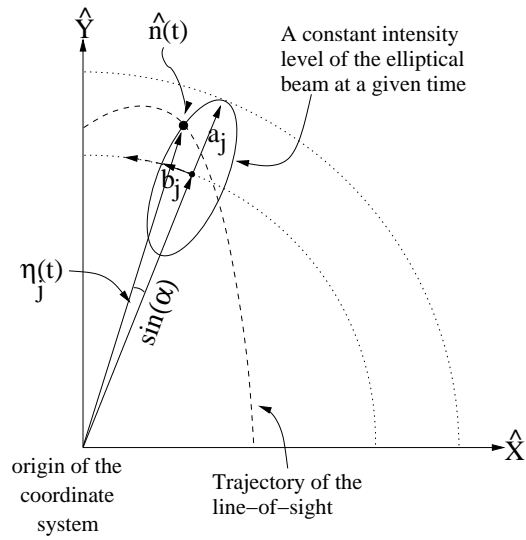}
\caption{
A zoomed-in view of the elliptical beam on the X-Y plane with a constant intensity level. The given constant intensity level crosses our line-of-sight when the line-of-sight is within the effective beam area (lightly shaded area in Fig~\ref{fig_dig}(b)). The semi-major and semi-minor axes, {\it a$_{j}$} and {\it b$_{j}$}, can take any value corresponding to the intensity level in a way such that the ratio a$_{j}$/b$_{j}$ is a constant. After projecting to the X-Y plane, the semi-major axis is aligned in the radially outward direction from the spin axis. $\sin$($\alpha$) is the distance to the center of the beam from the spin axis on the X-Y plane. {\it $\eta_{j}$(t)} is defined also on the X-Y plane, and it is the angle between the line joining the origin with the point at which the line-of-sight ($\hat{n}(t)$) encounters the constant intensity level of the beam and the line joining the spin axis with the center of the elliptical beam.
\label{fig_dig_new}}
\end{figure*}

\begin{figure*}
\epsscale{1.60}
\plotone{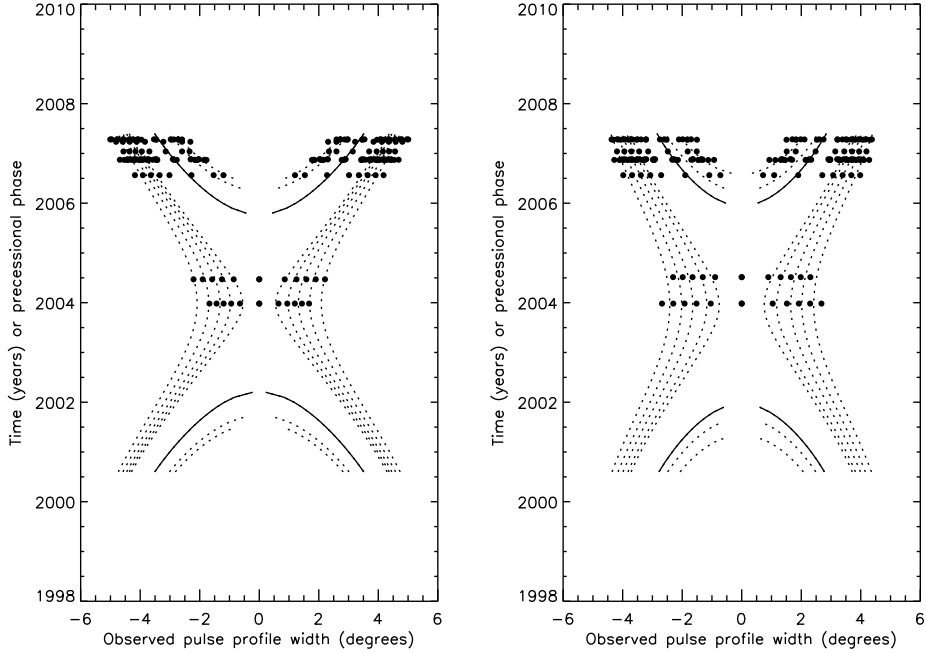}
\caption{
The two-dimensional pulse profile of PSR J0737$-$3039B, assuming a horse-shoe shaped beam. The best fit parameters for  BP1 are $\alpha$ $=$ 65.0$\degr$$_{-1.0^{\circ}}^{+2.0^{\circ}}$, $\theta$ $=$ 134.5$\degr$$_{-6.0^{\circ}}^{+1.5^{\circ}}$, and a/b $=$ 1.72$_{-0.20}^{+0.12}$ (left). For BP2, the best fit parameters are $\alpha$ $=$ 73.0$\degr$$_{-3.5^{\circ}}^{+1.4^{\circ}}$, $\theta$ $=$ 138.5$\degr$$_{-7.5^{\circ}}^{+4.0^{\circ}}$, and a/b $=$ 1.15$_{-0.22}^{+0.27}$ (right). We have omitted the data between MJDs $53175$ and $53939$. During this time, the profile was bi-modal but the separation between the peaks was small, making it difficult to measure widths at different intensity levels.
\label{mymodel}}
\end{figure*}

\begin{figure*}
\epsscale{1.60}
\plotone{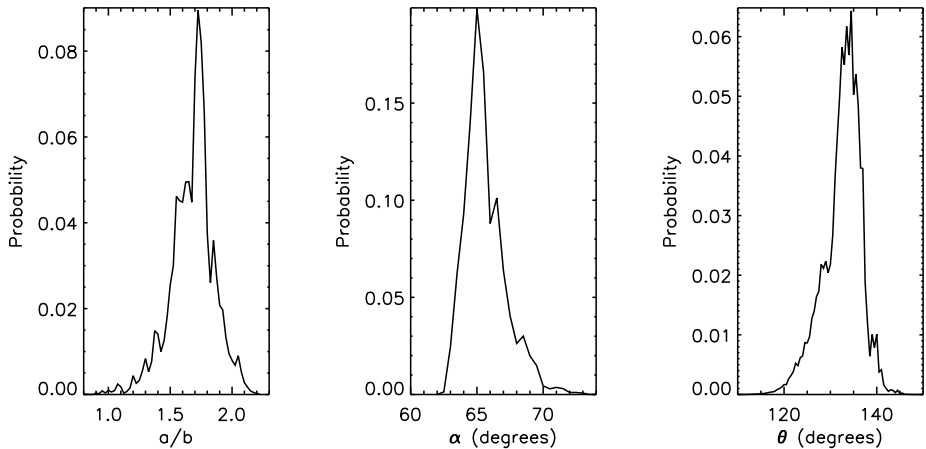}
\caption{The calculated posterior probabilities of the parameter values given by likelihood analysis for BP1. The best fit parameters and their 68\% confidence intervals are estimated to be a/b $=$ 1.72$_{-0.20}^{+0.12}$, $\alpha$ $=$ 65.0$\degr$$_{-1.0^{\circ}}^{+2.0^{\circ}}$ and $\theta$ $=$ 134.5$\degr$$_{-6.0^{\circ}}^{+1.5^{\circ}}$.
\label{like_error}}
\end{figure*}

\citet{lyu05} proposed a model to explain the change in the orbital longitudes of the two bright phases. It relies on the wind of pulsar A affecting the direction of emission of pulsar B and making it miss our line of sight for most of the orbit. In this model, the two bright phase regions evolve as a decreasing or increasing orbital phase gap between the two regions according to the direction of the spin precession \citep[see Fig. 6 of][]{lyu05}. However, the light curves of the two phases show that the real situation is somewhat different (see Fig.~\ref{shrink}). Motion of the ending orbital phase of BP1 to lower phases and starting orbital phase of BP2 to higher phases result in an increase in the orbital phase gap between the bright regions instead of the motion of both the starting and the ending phases that is predicted by the model. We are working on a more quantitative model in order to explain the evolution and the disappearance of B's emission.

\acknowledgments

We thank Joel Weisberg, Duncan Lorimer, and Maxim Lyutikov for very useful discussions. BBPP and MAM are supported by a WV EPSCoR grant. Pulsar research at UBC is supported by an NSERC DIscovery Grant.

\bibliographystyle{apj.bst}
\bibliography{psrrefs,modrefs,journals}

\end{document}